# Experimental generation of genuine four-partite entangled states with total three-party correlation for continuous variables

Aihong Tan    Yu Wang    Xiaoli Jin    Xiaolong Su

Xiaojun Jia   Jing Zhang   Changde Xie*   Kunchi Peng

(State Key Laboratory of Quantum Optics and Quantum Optics Devices, Institute of

Opto-Electronics, Shanxi University Taiyuan 030006, People's Republic of China )

We experimentally prepare a new type of continuous variable genuine four-partite entangled states, the quantum correlation property of which is different from that of the four-mode GHZ and cluster states, and which has not any qubit counterpart to be proposed at present. In the criterion inequalities for the full inseparability of the genuine four-partite entangled states, the amplitude and phase quadrature correlation variances totally consisting of three-party combination from the four entangled modes are involved. The measured correlation variances among the quadratures of the prepared entangled states satisfy the sufficient requirements for the full inseparability. The type of entangled states has especially potential application in quantum information with continuous quantum variables.

Entanglement is the most fascinating features of quantum mechanics and plays a central role in quantum information processing. In recent years, there has been an ongoing effort to characterize qualitatively and quantitatively the entanglement properties of multiparticle systems and apply them in quantum communication and information. The study of multipartite entangled states for qubit has shown that there exist different types of entanglement [1]. Most of the concepts of quantum information and computation have been initially developed for qubit, then were generalized in continuous variable (CV) field. The investigation of CV quantum information science has attracted extensive interest in recent years, due to its high efficiency in the generation, manipulation, and detection of the optical entangled states with continuous quantum variables [2]. The well-known genuine multipartite entangled states of the Greenberger-Horne-Zeilinger (GHZ) [3] and cluster states [4] were proposed firstly for qubits, then extended for CV [5,6]. A variety of quantum communication networks and quantum computation systems based on CV GHZ and CV cluster states of optical modes have been theoretically proposed [7-11]. The CV genuine three-partite and genuine four-partite GHZ and cluster entangled states of optical modes have been experimentally generated and successfully applied in the controlled dense coding quantum communication, quantum state sharing and quantum teleportation network [6, 12-15]. Very recently, a new type of CV genuine multipartite entangled states before its qubit counterpart was proposed, which is genuine maximum multipartite entangled states in the limit of infinite squeezing, and a novel nonlocal nondegenerate optical parametric amplifier (NOPA) and a telecloning scheme based on this entanglement were also designed [16,17]. In the criterion inequalities for detecting genuine CV GHZ and cluster entanglement, some two-party correlations of amplitude and phase quadratures of optical modes are included [16,18]. However, in each of the criterion inequalities for the new-proposed CV genuine four-partite entangled state only the correlation variances involving three-party combination of quadratures from the four entangled modes appear. Thus we named it the CV genuine four-partite entangled state with total three-party correlation (TTPC).

In this paper, we present the first experimental demonstration of the TTPC CV genuine four-partite entangled states. At first, we briefly deduce the criterion inequalitites for detecting the full inseparability of the entangled states, which are expressed with the experimentally measurable parameters. Then the experimental system and scheme are described. At last the experimental results are presented.

The principle schematic of the scheme generating the TTPC CV genuine four-partite entangled state of optical modes is drawn in Fig.1. For being compatible with the experimental system used by us and without losing generality, we assume that the EPR beams are produced from a pair of NOPA1 and NOPA2 operating in the state of de-amplification [6]. The pump fields and the injected signals ($a_{01} \sim a_{04}$) of both NOPAs come from a laser source for providing a pair of balanced EPR beams.

The input-output Heisenberg evolutions of the field modes of NOPA operating in the state of de-amplification are expressed by [19]

$$X_{a1(a3)} = X_{01(03)} \cosh r - X_{02(04)} \sinh r,$$

$$Y_{a1(a3)} = Y_{01(03)} \cosh r + Y_{02(04)} \sinh r,$$

$$X_{a2(a4)} = X_{02(04)} \cosh r - X_{01(03)} \sinh r, \qquad (1)$$

$$Y_{a2(a4)} = Y_{02(04)} \cosh r + Y_{01(03)} \sinh r,$$

where, $X_{a1(a3)}$, $X_{a2(a4)}$ ($Y_{a1(a3)}$, $Y_{a2(a4)}$) denote the amplitude (phase) quadratures of the output modes $a_1$ ($a_3$) and $a_2$ ($a_4$) of NOPA1 (NOPA2). $X_{01(03)}$, $X_{02(04)}$ ($Y_{01(03)}$, $Y_{02(04)}$) are the corresponding amplitude (phase) quadratures of the injected fields. In experiments, generally, the absolute values of all injected quadratures are made identical. $r(0 \leq r \leq \infty)$ is the squeezing factor which depends on the length and the effective second-order susceptibility of the nonlinear crystal used in NOPA, the losses of the optical cavity, as well as the intensity of pump field. We have assumed that the two squeezing factors for two NOPAs are the same for simplicity and the requirement is easy to be reached in the experiments because the two NOPAs were constructed in identical configuration, and the intracavity losses of the four modes were balanced.

From Eqs.(1), we can easily calculate the variances of the sum of amplitude quadratures and the difference of phase quadratures between mode $a_1$ ($a_3$) and $a_2$ ($a_4$):

$$\langle \delta^2(X_{a1(a3)} + X_{a2(a4)}) \rangle = \langle \delta^2(Y_{a1(a3)} - Y_{a2(a4)}) \rangle = \frac{1}{2} e^{-2r}, \qquad (2)$$

where the shot noise limit (SNL) of each submode of the entangled state has been normalized to $1/4$, and thus the SNL of total four submodes ($a_1 \sim a_4$) to 1.

Interfering modes $a_2$ and $a_3$ on a 50% beam-splitter (BS) and controlling the phase difference between them at $\pi/2$, the resultant four modes $b_1 \sim b_4$ equal to:

$$\begin{aligned}
b_1 &= a_1 = X_{a1} + iY_{a1}, \\
b_2 &= \frac{1}{\sqrt{2}}(a_2 + ia_3) = \frac{1}{\sqrt{2}}(X_{a2} - Y_{a3}) + \frac{i}{\sqrt{2}}(Y_{a2} + X_{a3}), \\
b_3 &= a_4 = X_{a4} + iY_{a4}, \\
b_4 &= \frac{1}{\sqrt{2}}(a_2 - ia_3) = \frac{1}{\sqrt{2}}(X_{a2} + Y_{a3}) + \frac{i}{\sqrt{2}}(Y_{a2} - X_{a3}).
\end{aligned} \quad (3)$$

In the ideal case corresponding to infinite squeezing ($r \to \infty$), the state is a simultaneous eigenstate of

$$\begin{aligned}
\sqrt{2}X_{b1} + X_{b4} + X_{b2} &\to 0, \\
\sqrt{2}Y_{b2} - Y_{b1} + X_{b3} &\to 0, \\
\sqrt{2}Y_{b3} + X_{b2} - X_{b4} &\to 0, \\
-\sqrt{2}Y_{b4} + X_{b3} + Y_{b1} &\to 0,
\end{aligned} \quad (4)$$

whose quantum correlations are different from CV GHZ and cluster state discussed in Ref. [5].

As well-known, the ideal case of $r \to \infty$ is non-physical and is not able to be reached in real experiments since infinite energy is required. Based on the same method of concluding the full inseparability criteria of multipartite CV GHZ entanglement in Ref. [18], we derived the sufficient requirements of full inseparability for the TTPC quadripartite entangled state. When the following three inequalities are satisfied simultaneously, the four submodes $b_i (i = 1 \sim 4)$ are in a fully inseparable entangled state:

$$\begin{aligned}
I\ & \left\langle \delta^2(\sqrt{2}X_{b2} + Y_{b3} + g_{x1}X_{b1}) \right\rangle + \left\langle \delta^2(Y_{b2} + \sqrt{2}X_{b3} - g_{y4}Y_{b4}) \right\rangle < \sqrt{2}, \\
II\ & \left\langle \delta^2(X_{b1} + Y_{b3} + \sqrt{2}g_{x2}X_{b2}) \right\rangle + \left\langle \delta^2(Y_{b1} + X_{b3} - \sqrt{2}g_{y4}Y_{b4}) \right\rangle < 1, \\
III\ & \left\langle \delta^2(X_{b2} + X_{b4} + \sqrt{2}g_{x1}X_{b1}) \right\rangle + \left\langle \delta^2(Y_{b2} - Y_{b4} + \sqrt{2}g_{x3}X_{b3}) \right\rangle < 1,
\end{aligned} \quad (5)$$

where, $g_i$ is the gain factor (arbitrary real parameter).

The correlation variances of quadratures of modes $b_i$ calculated from Eqs.(1) and (2) are

$$\langle \delta^2(\sqrt{2}X_{b2} + Y_{b3} + g_{x1}X_{b1}) \rangle = \frac{1}{2}[(g_{x1}-1)^2 e^{2r} + (g_{x1}+1)^2 e^{-2r} + e^{-2r}],$$

$$\langle \delta^2(Y_{b2} + \sqrt{2}X_{b3} - g_{y4}Y_{b4}) \rangle = \frac{1}{2}[(g_{y4}-1)^2 e^{2r} + (g_{y4}^2 + 2g_{y4} + 5)e^{-2r}],$$

$$\langle \delta^2(X_{b1} + Y_{b3} + \sqrt{2}g_{x2}X_{b2}) \rangle = (g_{x2}-1)^2 e^{2r} + (g_{x2}+1)^2 e^{-2r}, \quad (6)$$

$$\langle \delta^2(Y_{b1} + X_{b3} - \sqrt{2}g_{y4}Y_{b4}) \rangle = (g_{y4}-1)^2 e^{2r} + (g_{y4}+1)^2 e^{-2r},$$

$$\langle \delta^2(X_{b2} + X_{b4} + \sqrt{2}g_{x1}X_{b1}) \rangle = (g_{x1}-1)^2 e^{2r} + (g_{x1}+1)^2 e^{-2r},$$

$$\langle \delta^2(Y_{b2} - Y_{b4} + \sqrt{2}g_{x3}X_{b3}) \rangle = (g_{x3}-1)^2 e^{2r} + (g_{x3}+1)^2 e^{-2r}.$$

Calculating the minimum values of the left side expressions in Eqs. (5), we get the optimized gain factors

$$g_{x1}^{opt} = g_{x2}^{opt} = g_{x3}^{opt} = g_{y4}^{opt} = \frac{e^{4r}-1}{e^{4r}+1}. \quad (7)$$

In experiments the optimal gains can be met by adjusting the electronic gains of photocurrents to minimize the detected correlation variances.

Figure 2 shows the experimental setup. The laser is a homemade continuous wave intracavity frequency-doubled and frequency stabilized Nd-doped YAlO$_3$ perovskite (Nd:YAP)/KTP (KTiOPO$_4$, potassium titanyl phosphate) ring laser consisting of five mirrors. The second harmonic wave output at $0.54\,\mu m$ and the fundamental wave output at $1.08\,\mu m$ from the laser source are used for the pump fields and the injected signals of the two NOPAs (NOPA1 and NOPA2), respectively. For obtaining a pair of symmetric EPR entangled optical beams, the two NOPAs were constructed in identical configuration, both of which consist of an $\alpha$-cut type-II KTP crystal and a concave mirror. The front face of the KTP is coated to be used as the input coupler and the concave mirror as the output coupler of EPR beams is mounted on a piezo-electric transducer (PZT) for locking actively the cavity length of NOPA on resonance with the injected signal at $1.08\,\mu m$. Through a parametric down conversion process of type-II phase match, an EPR beam with anticorrelated amplitude quadratures and correlated phase quadratures may be produced from a NOPA operating in the state of de-amplification, that is, the pump field and the injected signal are out of phase [20]. The entangled two modes of EPR beams are just the signal and idler modes produced from the process, which have identical frequency with the injected signal at $1.08\,\mu m$ and the orthogonal polarization with each other [20]. They are split by the polarizing beam splitter PBS. Because the same laser serves as the pump field and the injected signal source of two NOPAs, the classical coherence between a pair of EPR beams generating from two NOPAs is ensured [21].

The phase difference between the interfering modes $a_2$ and $a_3$ on 50% BS is controlled at $\pi/2$ with a piezo-electric transducer and a feedback circuit. The four modes $b_i$ and the four local oscillation beams at $1.08\,\mu m$ deriving from the pump laser (not shown in Fig.2) are sent to the four sets of the balanced-homodyne detectors (BHD 1~4), respectively, for measuring the fluctuation variances of the amplitude or phase quadratures of mode $b_i$. The measured photocurrent variances of each mode are combined by the positive ($+$) or negative ($-$) power combiner according to the different requirements of the correlation variances in Eqs. (6), then are sent to a spectrum analyzer (SA) for recording the desired variety correlation variances.

The experimentally measured initial squeezing degree of the output fields from NOPA1 and NOPA2 equals to $2.6\pm 0.1$ dB below the SNL (the corresponding squeezing parameter $r$ equals to $0.30\pm 0.01$). During the measurements, the pump power of NOPAs at $0.54\,\mu m$ wavelength is about $110$ mW below the oscillation threshold of 150 mW and the intensity of the injected signal at $1.08\,\mu m$ is 9 mW.

Adjusting the electronic gains to the optimal values, the measured correlation variances of

$\left\langle \delta^2(\sqrt{2}X_{b2}+Y_{b3}+g_{x1}^{opt}X_{b1})\right\rangle, \left\langle \delta^2(Y_{b2}+\sqrt{2}X_{b3}-g_{y4}^{opt}Y_{b4})\right\rangle, \left\langle \delta^2(X_{b1}+Y_{b3}+\sqrt{2}g_{x2}^{opt}X_{b2})\right\rangle,$

$\left\langle \delta^2(Y_{b1}+X_{b3}-\sqrt{2}g_{y4}^{opt}Y_{b4})\right\rangle, \left\langle \delta^2(X_{b2}+X_{b4}+\sqrt{2}g_{x1}^{opt}X_{b1})\right\rangle, \left\langle \delta^2(Y_{b2}-Y_{b4}+\sqrt{2}g_{y3}^{opt}X_{b3})\right\rangle$

are $1.9\pm 0.1\,\text{dB}$, $1.2\pm 0.1\,\text{dB}$，$1.2\pm 0.1\,\text{dB}$，$0.7\pm 0.1\,\text{dB}$，$1.1\pm 0.1\,\text{dB}$，$0.5\pm 0.1\,\text{dB}$ below the SNL, respectively, which are shown in Fig.3. The optimal gains used in the measurements are $g^{opt}=0.41$.

Substituting the measured correlation variances into the left sides of Eq.(5), we have $I=1.11\pm 0.01$，$II=0.94\pm 0.01$，$III=0.97\pm 0.01$. All of them are smaller than the normalized SNL. It means that the obtained modes are in a fully inseparable TTPC entangled state.

For achieving the measurements of a variety of the correlation variances under same experimental conditions, the pump laser has to be operated stably in the whole experimental process. The resonating frequencies of the laser and the two NOPAs must be locked in a longer term. The standard sideband frequency locking and the interference feedback technologies are utilized for the frequency and the relative phase locking respectively.

For conclusion, we experimentally produced a new type of CV genuine four-partite entangled states, which represents a nice example where a CV quantum information protocol is obtained before its qubits counterpart. By means of the interference of two modes deriving from two EPR entangled states on an optical beam-splitter, the quantum entanglement among the four modes is established. The TTPC CV entangled states may be applied in the nonlocal NOPA and CV quantum information systems in which the total three-party correlation is required. It is significant to understand the special features of entanglement appearing in states of infinite-dimensional Hilbert spaces at first, and then to push the development of qubit entanglement.

This research was supported by the PCSIRT (Grant No. IRT0516), Natural Science Foundation of China (Grants No. 60608012, 60736040 and 10674088)，NSFC for Distinguished Young Scholars(Grant No.10725416), National Basic Research Program of China(Grant No.2006CB921101).

\* Corresponding author. Email address: changde@sxu.edu.cn

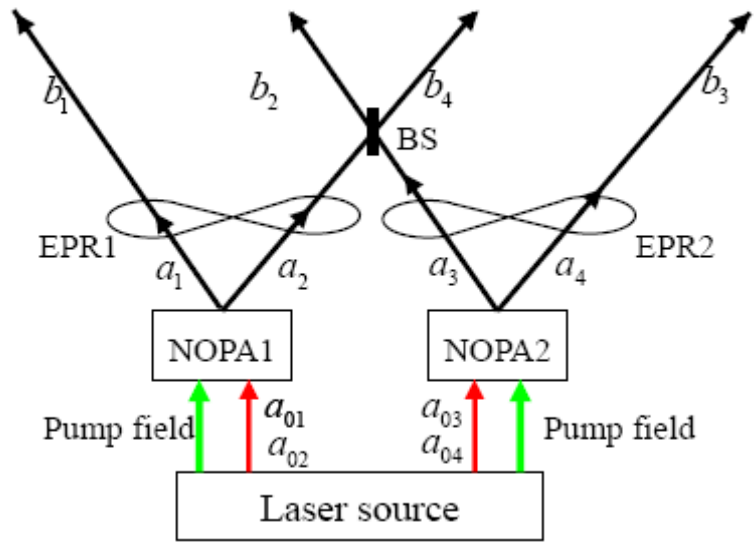

Fig.1 Principle schematic for CV quadripartite TTPC generation

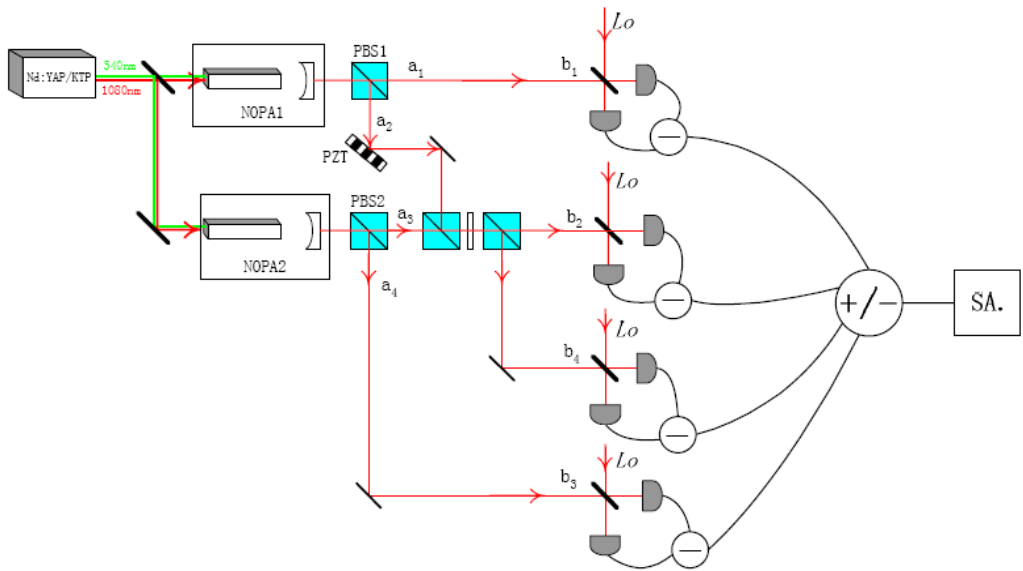

Fig.2    Schematic of the experimental setup for the TTPC generation

Nd:YAP/KTP---laser source; NOPA---nondegenerate optical parametric amplification;
PBS---polarizing optical beamsplitter; PZT---piezoelectric transducer;
+/- ---positive/negative power combiner; SA---spectrum analyzer.

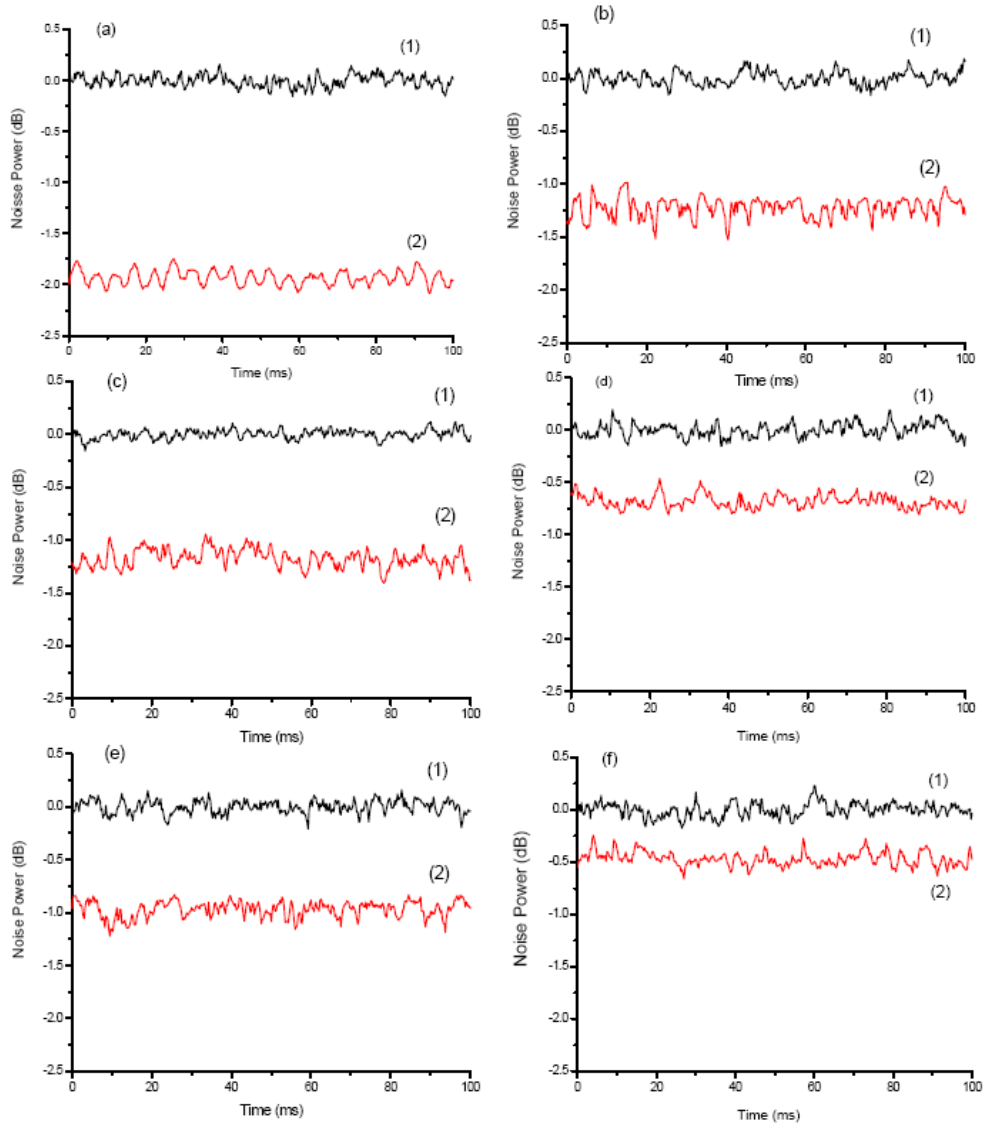

Fig.3 The measured correlation variances of the TTPC at 2 MHz

(a) $\left\langle \delta^2(\sqrt{2}X_{b2} + Y_{b3} + g_{x1}^{opt}X_{b1}) \right\rangle$ (b) $\left\langle \delta^2(Y_{b2} + \sqrt{2}X_{b3} - g_{y4}^{opt}Y_{b4}) \right\rangle$ (c) $\left\langle \delta^2(X_{b1} + Y_{b3} + \sqrt{2}g_{x2}^{opt}X_{b2}) \right\rangle$

(d) $\left\langle \delta^2(Y_{b1} + X_{b3} - \sqrt{2}g_{y4}^{opt}Y_{b4}) \right\rangle$ (e) $\left\langle \delta^2(X_{b2} + X_{b4} + \sqrt{2}g_{x1}^{opt}X_{b1}) \right\rangle$ (f) $\left\langle \delta^2(Y_{b2} - Y_{b4} + \sqrt{2}g_{x3}^{opt}X_{b3}) \right\rangle$

(1) The corresponding shot noise limit. (2) The correlation noise power.
The measurement parameters of SA: RBW(resolution bandwith)—30 kHz;
VBW(video bandwith)—30 Hz